\documentclass{ifacconf}
\usepackage{graphicx}
\usepackage{natbib}
\usepackage[hyphens]{url}     
\usepackage{xcolor}
\usepackage{float}
\usepackage{amsmath}
\usepackage{subcaption}
\usepackage{dblfloatfix}
\usepackage{enumitem}

\newcommand{\dg}{^\circ}
\begin{document}

\begin{frontmatter}

\title{Mitigating Dynamic Tip-Over during Mobile Crane Slewing using Input Shaping }

\author[First]{Navneet Kaur}
\author[Second]{Christopher J. Adams}
\author[Second]{William E. Singhose}
\author[First]{Santosh Devasia}

\address[First]{University of Washington, 
   Seattle, WA 98195 USA,  
 \\ (e-mails: navneet@uw.edu, devasia@uw.edu).
   }
\address[Second]{Georgia Institute of Technology, GA 30332 USA, \\
(e-mails: cjadams@gatech.edu, singhose@gatech.edu).
}

\begin{abstract}
Payload swing during rapid slewing of mobile cranes poses a safety risk, as it generates overturning moments that can lead to tip-over accidents of mobile cranes. Currently, to limit the risk of tip-over, mobile crane operators are forced to either reduce the slewing speed (which lowers productivity) or reduce the load being carried to reduce the induced moments. Both of these approaches reduce productivity. This paper seeks to enable rapid slewing without compromising safety by applying input shaping to the crane-slewing commands generated by the operator. A key advantage of this approach is that the input shaper requires only the information about the rope length, and does not require detailed mobile crane dynamics. Simulations and experiments show that the proposed method reduces residual payload swing and enables significantly higher slewing speeds without tip over, reducing slewing completion time by at least $38\%$ compared to unshaped control. Human control with input shaping improves task completion time by $13\%$, reduces the peak swing by $18\%$, and reduces the potential of collisions by $82\%$ when compared to unshaped control. Moreover, shaped control with a human had no tip-over, whereas large swing led to tip-over without input shaping. Thereby, the proposed method substantially recovers the operational-safety envelope of mobile cranes (designed to avoid tip-over using static analysis) that would otherwise be lost in dynamic conditions. Videos and demonstrations are available at \url{https://youtu.be/dVy3bbIhrBU}. 
\end{abstract}


\end{frontmatter}

\section{Introduction}
\begin{figure}[htbp]
    \centering
    \includegraphics[width=7.5cm]{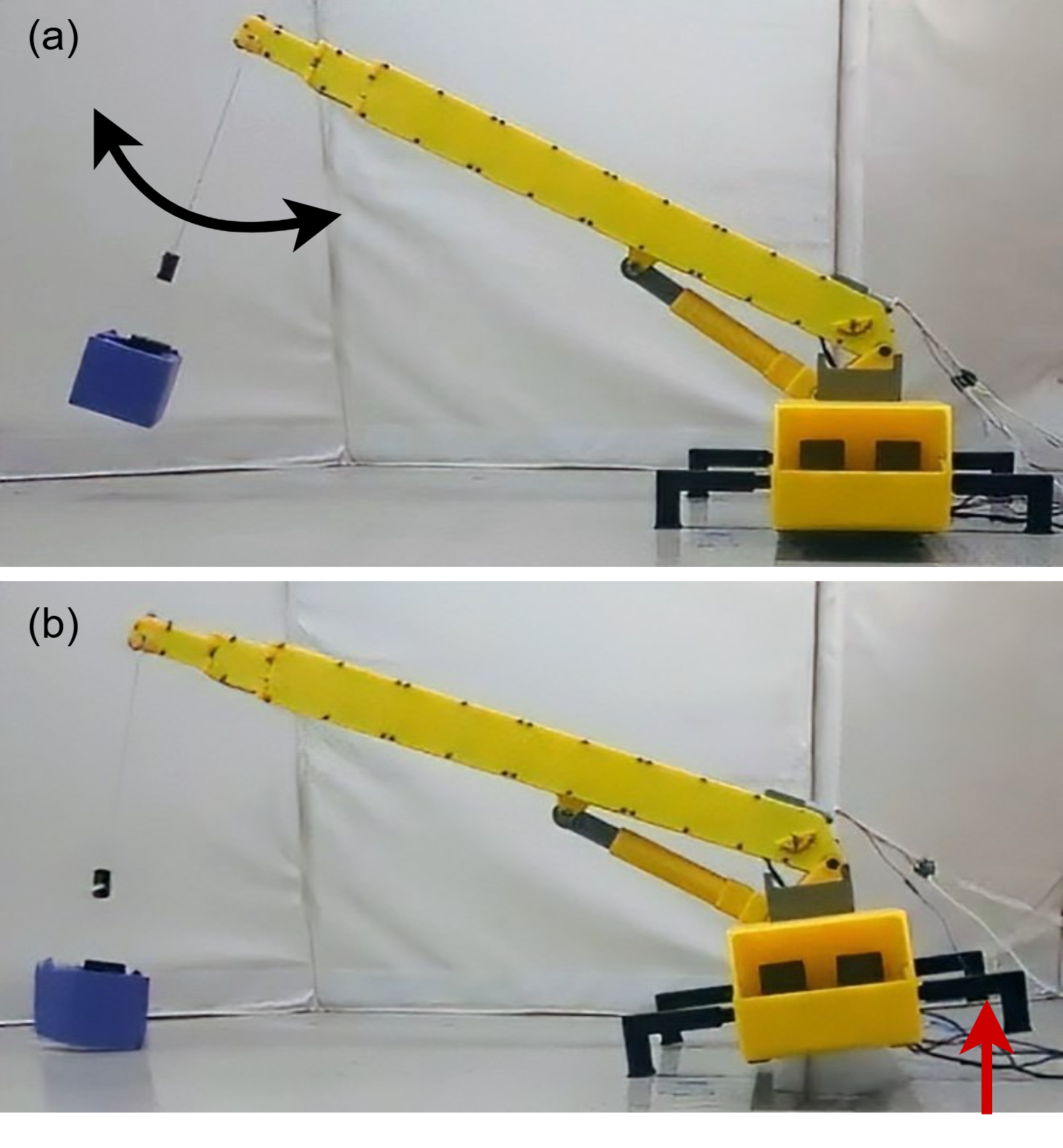}
    \caption{Experimental demonstration of mobile crane tip-over instability due to dynamics. (a) Payload swings due to slewing motion. (b) Outriggers of crane lift off at the onset of tip-over. }
    \label{fig_overview}
\end{figure}
Mobile cranes are widely utilized in construction, shipping, and industrial operations, where both safety and productivity are crucial. Despite their versatility, mobile cranes remain highly susceptible to instability, with tip-over accidents continuing to be a major cause of fatalities and equipment damage (\cite{Tip_Over_Stability_Analysis, tipover_stability_analysis_mobile, noish_alert, OSHA_MobileCraneAccidents, tipover_responses, stability_analysis}). A significant contributor to such incidents is the dynamic interaction between the boom and the suspended payload during slewing motion. In practice, load limits for crane safety are prescribed by load charts that are based on static equilibrium analysis (\cite{GroveGMK7550, mobile_crane_inspector}) to specify the maximum allowable load as a function of crane configuration. These limits do not account for dynamic effects such as payload swing due to slewing. As a result,  crane operations that are theoretically within the static safe limits may still become unstable due to dynamic effects, as shown in Fig.~\ref{fig_overview}. The goal of this work is to avoid tip-over due to excessive dynamics-induced payload swing during slewing, even with statically safe loads. 

Tip-over can be mitigated by modifying the crane’s mechanical configuration or by automating its motion. For example, increasing the counterweight mass has been shown to expand the stability margin and improve lifting capacity in crawler cranes (\cite{lifting_path_planning}). Other approaches focus on controlling outrigger forces to stabilize the crane during motion (\cite{synchronous_control_theory}), though moving outriggers during an operation is generally unsafe. Additional studies examine specific crane architectures and derive model-based controllers for swing suppression (\cite{modeling_and_observer_based}). While these works demonstrate methods for improving crane stability, most rely on precise mechanical actuation, detailed system models, or structural modifications that are impractical for typical mobile crane operations. Prior work has shown that FIR-based filtering can reduce sway during crane motion, but it does not address dynamic stability envelopes for tip-over  (\cite{digital_filter}). Tip-over can be avoided by reducing payload swing through automated trajectory planning when accurate models are available and environmental conditions are fully known. Recent robotic crane systems employ fully autonomous path planning, obstacle avoidance, and motion compensation to achieve reliable point-to-point load transport (\cite{robotic_tower_cranes, lifting_path_planning, optimal_traj_planning}). These autonomous systems operate without continuous human involvement, relying on pre-planned trajectories. However, fully removing human control overlooks the operator’s intuitive judgment, which is essential in unstructured and dynamic construction environments. Therefore, there is a need for approaches that reduce swing-induced tip-over risk by augmenting human control instead of replacing it. In this work, control techniques are applied to assist manual slewing operations by reducing swing-induced forces while maintaining the operator’s real-time decision authority.

The main contribution of this work is that it mitigates the tip-over of mobile cranes caused by dynamic effects. The method shapes the operator’s motion commands (the input) to cancel vibrations at the system’s natural frequency of the swing dynamics. Specifically, a Modified Unity-Magnitude Zero-Vibration (MUMZV) input shaper (\cite{mumzv}) is applied to the slewing velocity command to suppress the swing mode. A major advantage of the proposed approach (using input shaping to reduce payload swing) is that it does not require a detailed crane model.  Since the dominant swing frequency depends primarily on the rope length, the input-shaping filter can be designed directly based on this parameter, making it effectively model-free and straightforward to implement. By ensuring that changes in slewing speed do not excite the natural frequencies of the suspended payload, the shaped command minimizes residual oscillations and keeps the motion within safe bounds. Consequently, the MUMZV-based human-in-the-loop approach enables higher slewing speeds while avoiding tip-over. The simulation and experimental results show that the proposed approach substantially recovers the operational-safety envelope of mobile cranes (designed using static analysis) that would otherwise be lost in dynamic conditions.

The paper is organized as follows. Section II formulates the problem, introducing the mobile crane slewing dynamics and the associated tip-over risks. Section III presents the methodology, including the design of the MUMZV input shaper used to suppress payload oscillations during manual slewing. Section IV reports the simulation and experimental results, demonstrating reductions in swing and increases in safe slewing speed. Section V concludes the paper and discusses directions for future work.

\section{Problem Formulation}
Tip-over risk increases during slewing because payload swing shifts the payload outward, creating extra overturning moment beyond what the static configuration predicts. When this additional moment is present, load–boom configurations that are stable under the static load conditions can become unstable once slewing begins.

\begin{figure}[tb]
    \centering
    \includegraphics[width=8.4cm]{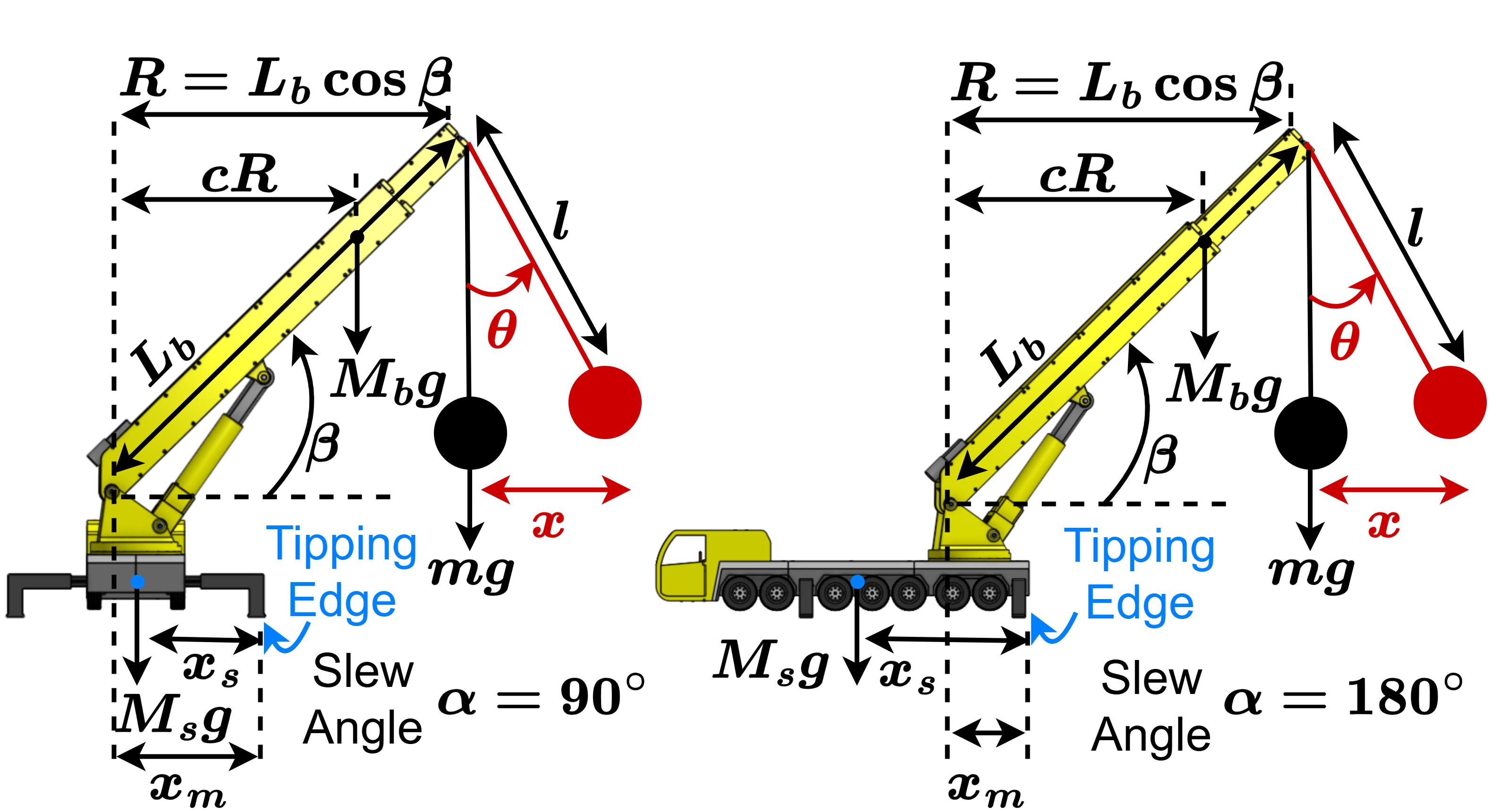}
    \caption{Schematic illustrating the overturning moment contributions for slewing at slew angles $\alpha = 90\dg$ and $\alpha = 180\dg$. }
    \label{fig_moment_explain}
\end{figure}

\subsection{Effect of Payload Swing on Overturning Moments}
The static tipping condition is based on balancing the gravitational moments about the relevant outrigger edge, as shown in Fig.~\ref{fig_moment_explain}, i.e., 
\begin{equation}
\label{eq_moment_balance_simple}
M_s g x_s \ge M_b g (cR - x_m) + m g (R - x_m) +
\boxed{\,mgx}
\end{equation}
with three major lumped masses: the lower structure of mass $M_s$, the boom of mass $M_b$, and the payload of mass $m$. For a boom of length $L_b$ at elevation angle $\beta$, the horizontal reach is $R = L_b \cos\beta$, which is referred to as the boom radius. The boom center of mass (CoM) lies at $cR$ from the slewing axis, where $c$ is a constant depending on the boom composition. $x_m$ denotes the horizontal distance from the boom base to the tipping edge, while $x_s$ denotes the distance from the lower-structure CoM to the same tipping edge. The payload normally acts at a distance $R$, however, when it swings out by an angle $\theta$, its horizontal position shifts outward by an amount $x = l \sin(\theta)$, where $l$ is the rope length, as shown in Fig.~\ref{fig_moment_explain}.

The overturning moment $mgx$, produced by the payload swing, is an additional destabilizing moment when compared to static conditions. The main structure mass $M_s$ (which often includes additional counterweights) contributes a stabilizing moment that needs to overcome the moments generated by the boom mass  $M_b$ and the load mass $m$, which act in the overturning direction. However, when the payload swings outward by angle $\theta$, the distance $x$ increases, increasing the destabilizing right-hand side of \eqref{eq_moment_balance_simple} and thereby reducing the stability margin. Even small swing oscillations can turn a previously stable configuration into an unstable one, especially when the direction of swing aligns with the direction of maximum potential for overturning such as slew angles $\alpha = 90\dg$ (when the main mass $M_s$ is closest to the tipping-outrigger edge) and $\alpha = 180\dg$ (when the boom's CoM is farthest away from the tipping-outrigger edge).

\subsection{Static Stability and Load Chart Analysis}
\begin{figure}[tb]
    \centering
    \includegraphics[width=8.4cm]{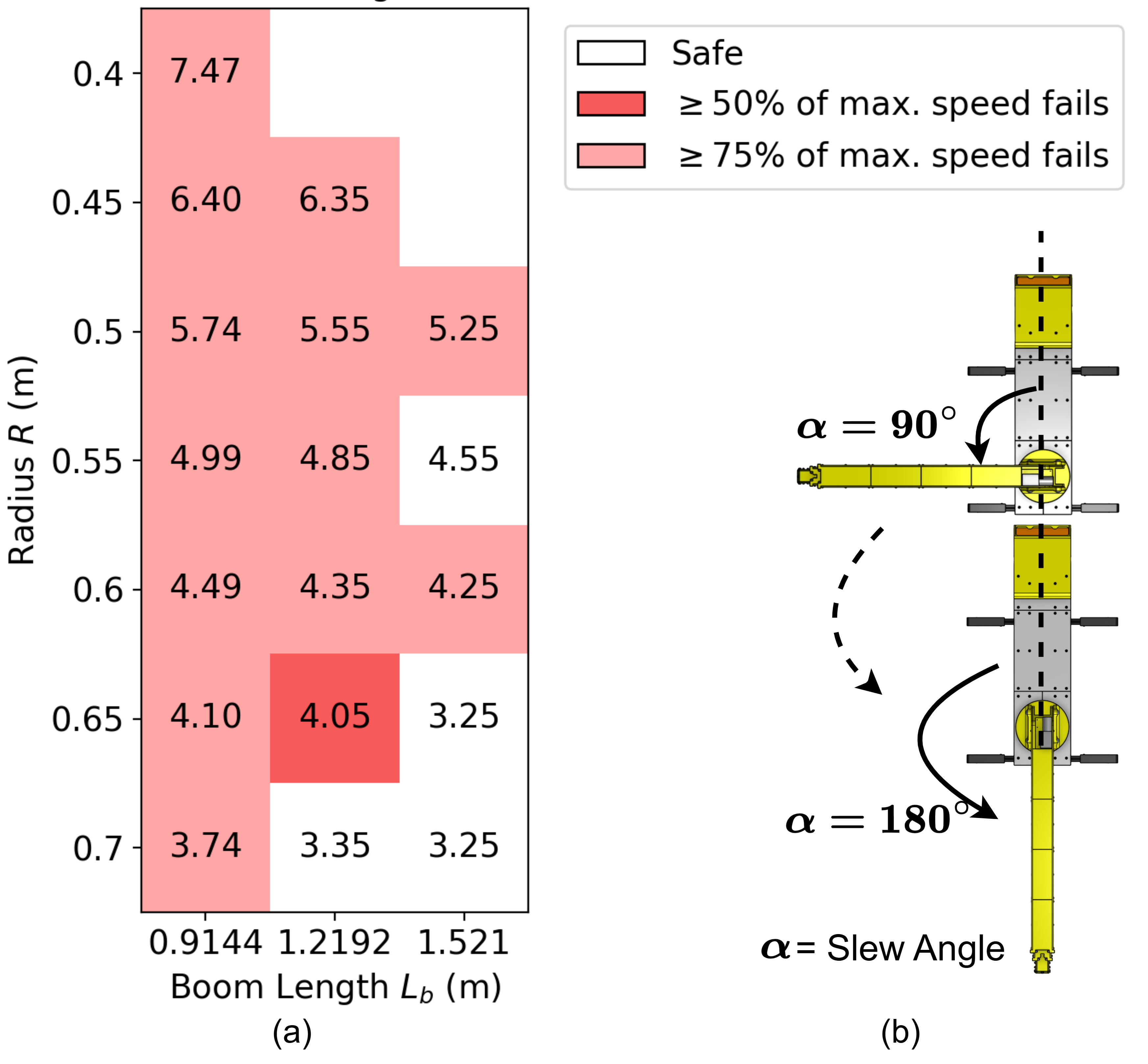}
    \caption{Reduction of safety envelope due to slewing dynamics (a) Load chart showing maximum statically-safe load.
    (b) Top-view of slewing motion used for each configuration.}
    \label{fig_loadchart}
\end{figure}

Load charts are used to inform crane operators about the maximum payload that the crane can support for each combination of boom radius $R = L_b \cos\beta$ and boom length $L_b$. Fig.~\ref{fig_loadchart}(a) shows the resulting chart for the scaled crane model, which was generated using the CoM positions and the base tipping axis for combinations of boom radius and boom length. This chart defines the maximum payload that can be supported without violating static stability or exceeding the allowable bending moment at the turntable.

The maximum speed ${\dot{\alpha}}$ was selected to maintain dynamic similarity between the scaled-model crane and a full-size crane. For full-size cranes, a typical maximum slewing speed $\dot{\alpha}$ is approximately 1.1 rpm ($6.6\dg/s$), with payload-swing natural frequency of $\omega_n=0.908$ Hz for a rope length $l = 11.89$ m (corresponds to $39'$).  To preserve the ratio $\dot{\alpha}/\omega_n$ between task frequency and swing frequency, the corresponding maximum slewing speed for dynamic similarity is about $32.51\dg/s$ when  the natural frequency for the scaled model is $\omega_n=4.47$ Hz. Using the static load chart as a baseline, each boom radius and length pair $(R, L_b)$ was then evaluated dynamically by commanding a $90\dg$ slewing motion at constant speeds $\dot{\alpha}$ of $50\%$, $75\%$, and $100\%$ of the maximum slewing speed calculated. For each configuration, the payload corresponding to the static limit was applied. The simulation recorded whether the crane remained stable or tipped during the motion. The failure map in Fig.~\ref{fig_loadchart}(a) illustrates that several configurations that are stable statically become unstable dynamically due to swing-induced increases in overturning moment.

\begin{figure}[t]
    \centering
    \includegraphics[width=8.4cm]{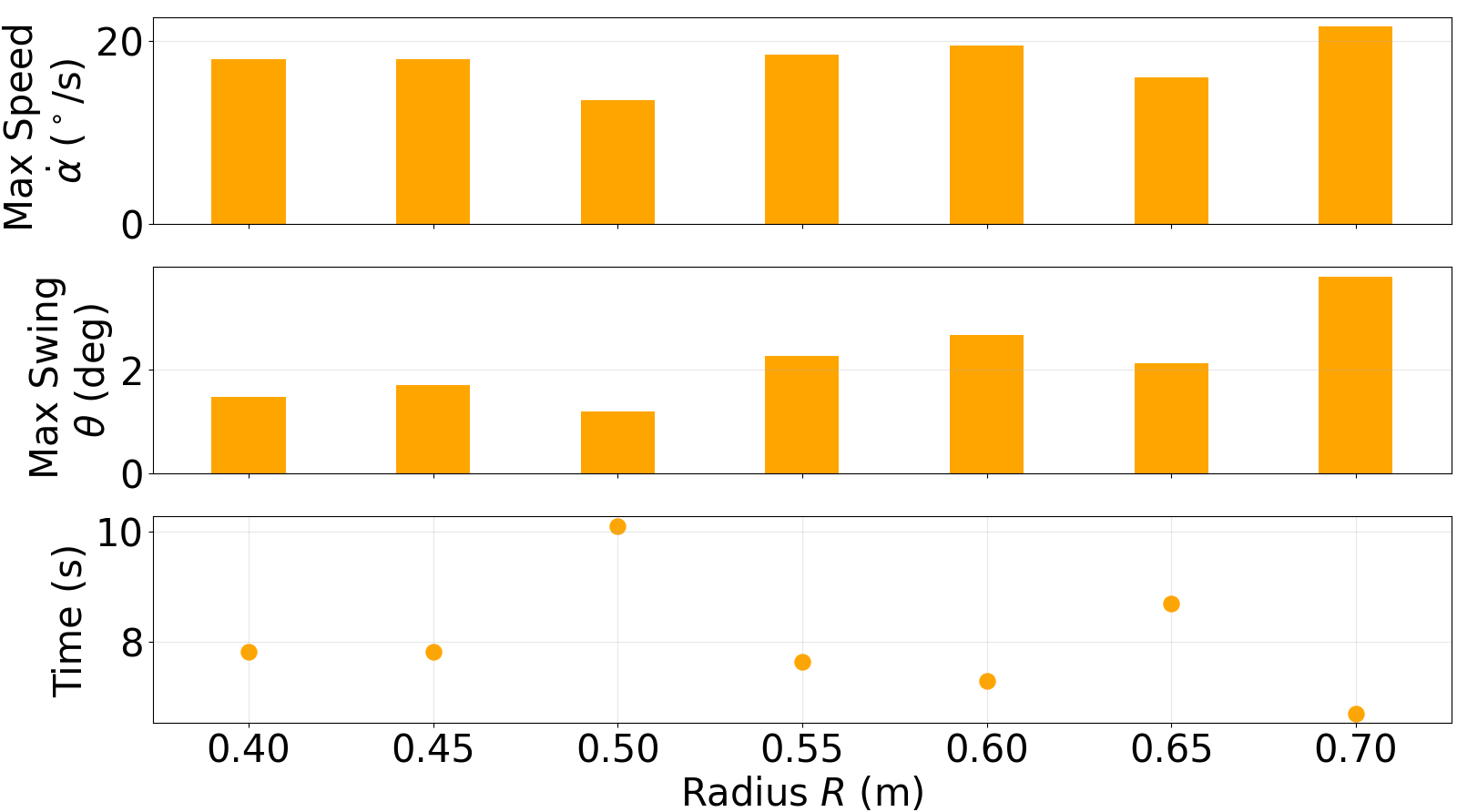}
    \caption{Dynamic limits for a fixed boom length $L_b = 0.9144$~m and maximum load $m$ (from  Fig.~\ref{fig_loadchart}) for different boom radius $R$: maximum safe slewing speed (top), corresponding peak swing angle (middle), and time to complete a $90\dg$ rotation (bottom).}
    \label{fig_speed_analysis}
\end{figure}

Although the crane may be statically stable, slewing can induce payload swing that can reduce the stability margin and cause tip-over. Fig.~\ref{fig_speed_analysis} summarizes the dynamic limits for a fixed boom length ($L_b = 0.9144$ m) by evaluating the maximum safe slewing speed, the corresponding peak payload swing, and the time required to complete a $90\dg$ rotation for the boom radius and maximum payload mass $(R,m)$ pair from Fig.~\ref{fig_loadchart}. The top plot reports the highest slewing speed at which the crane remained stable for that radius. Speeds above this limit resulted in crane tip-over. The middle plot shows the maximum swing angle $\theta_{\text{max}}$ recorded during the motion, which directly influences the additional overturning moment term $mgx$ in the stability condition. The bottom plot shows the total maneuver time corresponding to the highest slewing speed, which increases when the slewing speed is reduced. These trends show that even configurations that are statically admissible can become dynamically unstable due to swing-induced moments, and hence, there is a need for a method that increases the safe operating range under dynamic conditions.

The objective of this work is to mitigate swing-induced instability during slewing, enabling higher slewing speeds while maintaining safe stability margins and thereby recovering some of the statically-safe operating envelope lost due to dynamics.

\section{Methodology}

Payload swing behaves as a pendulum that is directly forced by the crane’s slewing acceleration (\cite{dynamics_and_control}). Applying the Euler–Lagrange equations to the crane–pendulum model gives the swing dynamics (\cite{command_shaping_boom_crane}) as,
\begin{equation}
    \begin{aligned}
        l\ddot{\theta}_1 + 2l\dot{\alpha}\dot{\theta}_2 + (g-l\dot{\alpha}^2)\theta_1 + l\ddot{\alpha}\theta_2 + L_b\ddot{\alpha}\cos\beta &= 0\\
        l\ddot{\theta}_2 - 2l\dot{\alpha}\dot{\theta}_1 + (g - l\dot{\alpha}^2)\theta_2 - l\ddot{\alpha}\theta_1 - L_b\dot{\alpha}^2\cos\beta &= 0
    \end{aligned}
    \label{eq_eom}
\end{equation}
where $\theta_1$ and $\theta_2$ are tangential and radial components of the swing angle, respectively. $l$ is the length of rope, $L_b$ is the boom lenght, $\alpha$ is the slew angle, and $\beta$ is the boom elevation angle. The natural frequency of this system would be
\begin{equation}
\omega_n = \sqrt{\frac{g}{l}}.
\label{eq_omega_n}
\end{equation}
Hence, slewing accelerations that contain components near the natural frequency $\omega_n$ will resonate with the swing dynamics mode and increase payload swing.

\subsection{Acceleration–Domain MUMZV Input Shaping}
Payload swing is driven by the slewing acceleration $\ddot{\alpha}(t)$ input, so input shaping is applied in the acceleration domain to not resonate the swing dynamics. Let $r(t)$ denote the operator’s desired slewing acceleration. A MUMZV shaper (\cite{mumzv}) is defined by three impulse time instants $t_1$, $t_2$, $t_3$ with corresponding amplitudes $A_1$, $A_2$, $A_3$ that satisfy
\begin{equation}
\sum_{i=1}^3 A_i e^{j\omega_n t_i} = 0,
\label{eq_mumzv_constraints}
\end{equation}
zero residual swing at $\omega_n$. The complete set of impulse times and amplitudes is determined by the deflection ratio $D_L$, which sets the fraction of the peak allowable swing. For a given $D_L \in (0, 0.5]$, the MUMZV timing formulas yield the time instants $t_2$, $t_3$, and the corresponding amplitude $A_3=2D_L$, with the mode period given by $T=2\pi/\omega_n$. The continuous-time shaper is defined by the function 
\begin{equation}
    s(t)=\sum_iA_i\delta(t-t_i),
\end{equation}
and the shaped acceleration command $u(t)$ for a time-varying reference input $r$ from a joystick  can be  obtained by convolution,
\begin{equation}
u(t) = (r * s)(t)
= \sum_i A_i, r(t - t_i).
\label{eq_conv}
\end{equation}

\subsection{From Shaped Acceleration to Slewing Velocity}
\begin{figure}[tb]
    \centering
    \includegraphics[width=8.4cm]{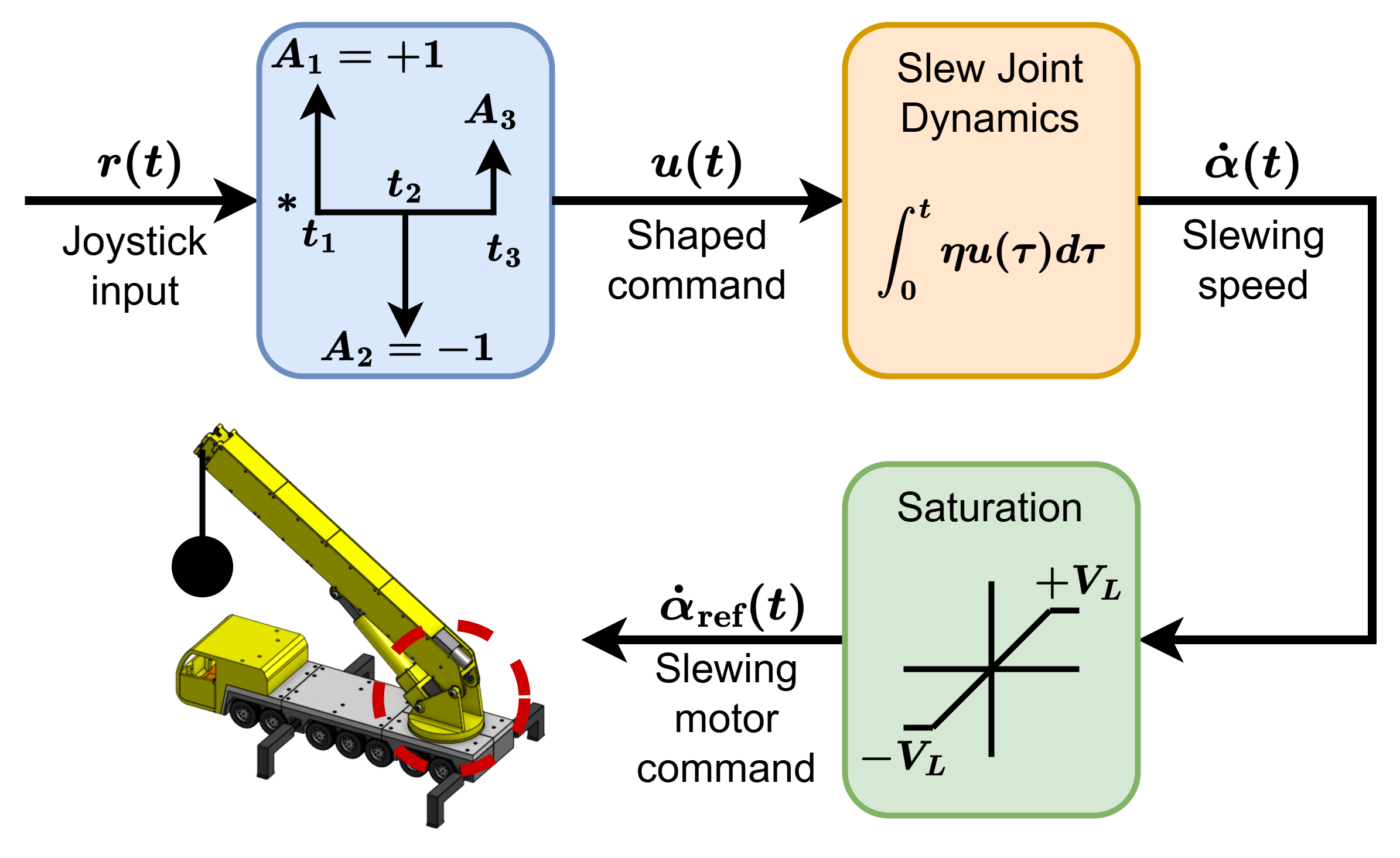}
    \caption{Block diagram of the MUMZV-based input shaping framework.}
    \label{fig_mumzv_block}
\end{figure}

The shaped slewing motion seeks to reduce residual payload swing at the natural frequency $\omega_n$ while maintaining bounded speed and input torque. 
The shaped command $u(t)$ is mapped to physical slewing acceleration via

\begin{equation}
    \ddot{\alpha}(t) = \eta\,u(t),
    \qquad 
    \eta = \frac{\tau_{\max}}{J_{\mathrm{eq}}},
    \label{eq_acc}
\end{equation}
where $\tau_{\max}$ is the maximum available amplitude of the slewing torque and $J_{\mathrm{eq}}$ is the equivalent inertia. The desired slewing speed profile is determined by integrating~\eqref{eq_acc} as,
\begin{equation}
    \dot{\alpha}_{\mathrm{ref}}(t)
    = \mathrm{sat}\!\left(
        \dot{\alpha}(0) 
        + \eta\int_{0}^{t} u(\tau)\,d\tau,\;
        -V_L,\; V_L
    \right),
    \label{eq_vref}
\end{equation}
where $V_L$ is the allowed speed limit and  $\mathrm{sat}(\cdot)$ denotes the saturation function that limits the argument to lie within the bounds $[-V_L, V_L]$ to limit the maximum slewing speed while preserving the MUMZV swing–cancellation property. This desired slewing speed $\dot{\alpha}_{\mathrm{ref}}$ is send as a reference speed command to the motor-speed controller for generating the slewing motion. The overall MUMZV input-shaping pipeline, from joystick input $r$  to the commanded slewing velocity $\dot{\alpha}_{\mathrm{ref}}$, is shown in Fig.~\ref{fig_mumzv_block}.

\section{Results}
\subsection{Simulation Validation}
\subsubsection{Simulation Setup}
\begin{figure}[b]
    \centering
    \includegraphics[width=8.4cm]{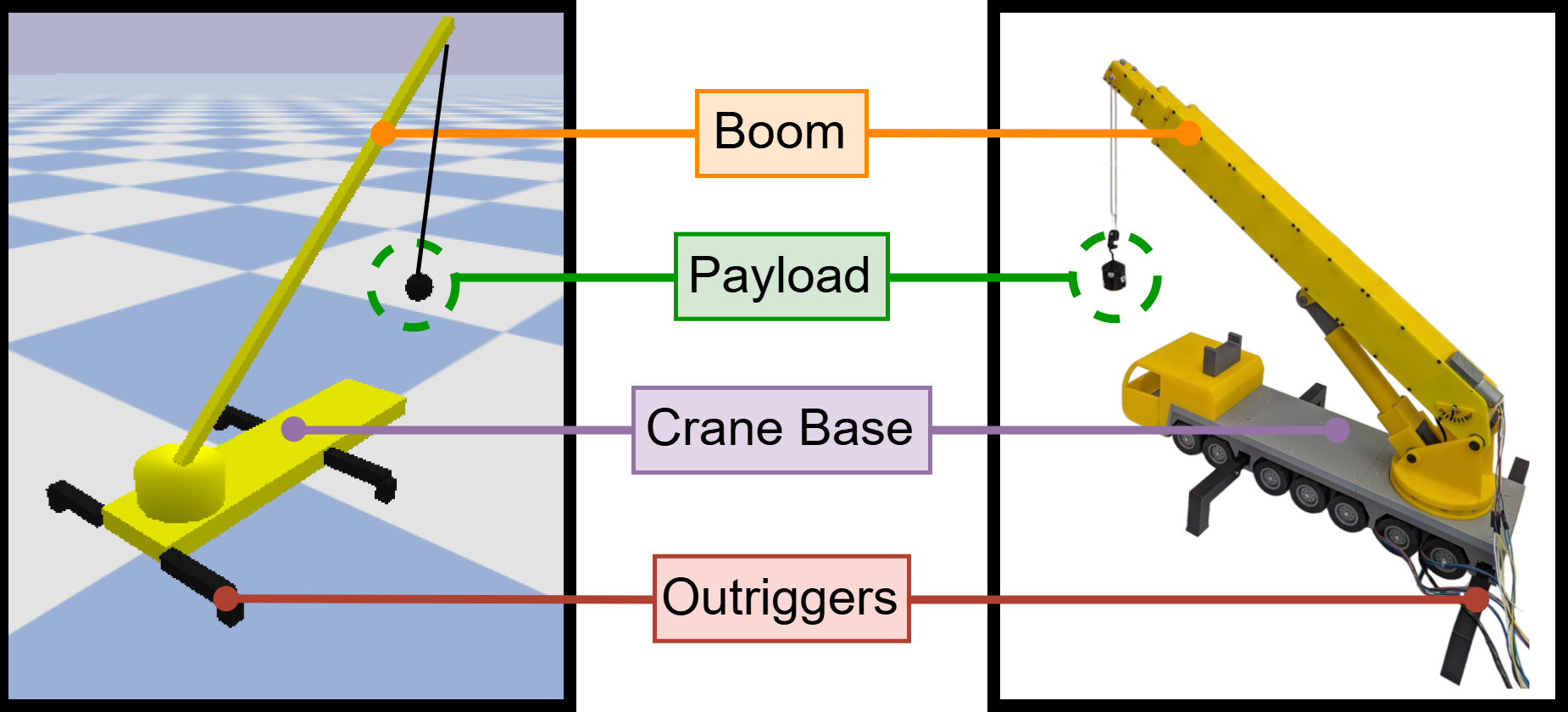}
    \caption{Scaled mobile crane model: simulation setup (left) and experimental setup (right). Identical geometry and mass enable direct comparison of swing behavior, tip-over stability limits, and achievable slewing speeds between simulation and physical tests. }
    \label{fig_sim_real_model}
\end{figure}

The simulation environment models the full crane system in PyBullet, including the base, turntable, boom, outriggers, and suspended payload. The physical parameters match the laboratory crane prototype with boom mass $M_b=2.93\;\mathrm{kg}$ and lower structure mass $M_s=7.24\;\mathrm{kg}$. Gravity is set to $g=-9.81\;\mathrm{m/s^2}$. Fig.~\ref{fig_sim_real_model} shows the PyBullet model alongside the experiment model crane.

The simulation time step was $1/240$ s and PyBullet's velocity-control mode was used with the shaped command $\dot{\alpha}_{\mathrm{ref}}(t)$. The MUMZV shaper was constructed from the rope length using~\eqref{eq_omega_n}–\eqref{eq_vref} with a deflection ratio $D_L=0.3$, yielding an amplitude $A_3=0.6$. The resulting impulse times were extended in $T$-spaced segments to form the three-level sequence, and the overall duration was adjusted to satisfy the slewing-speed limit. 

\subsubsection{Simulation Results}

\begin{figure}[t]
    \centering
    \includegraphics[width=8.4cm]{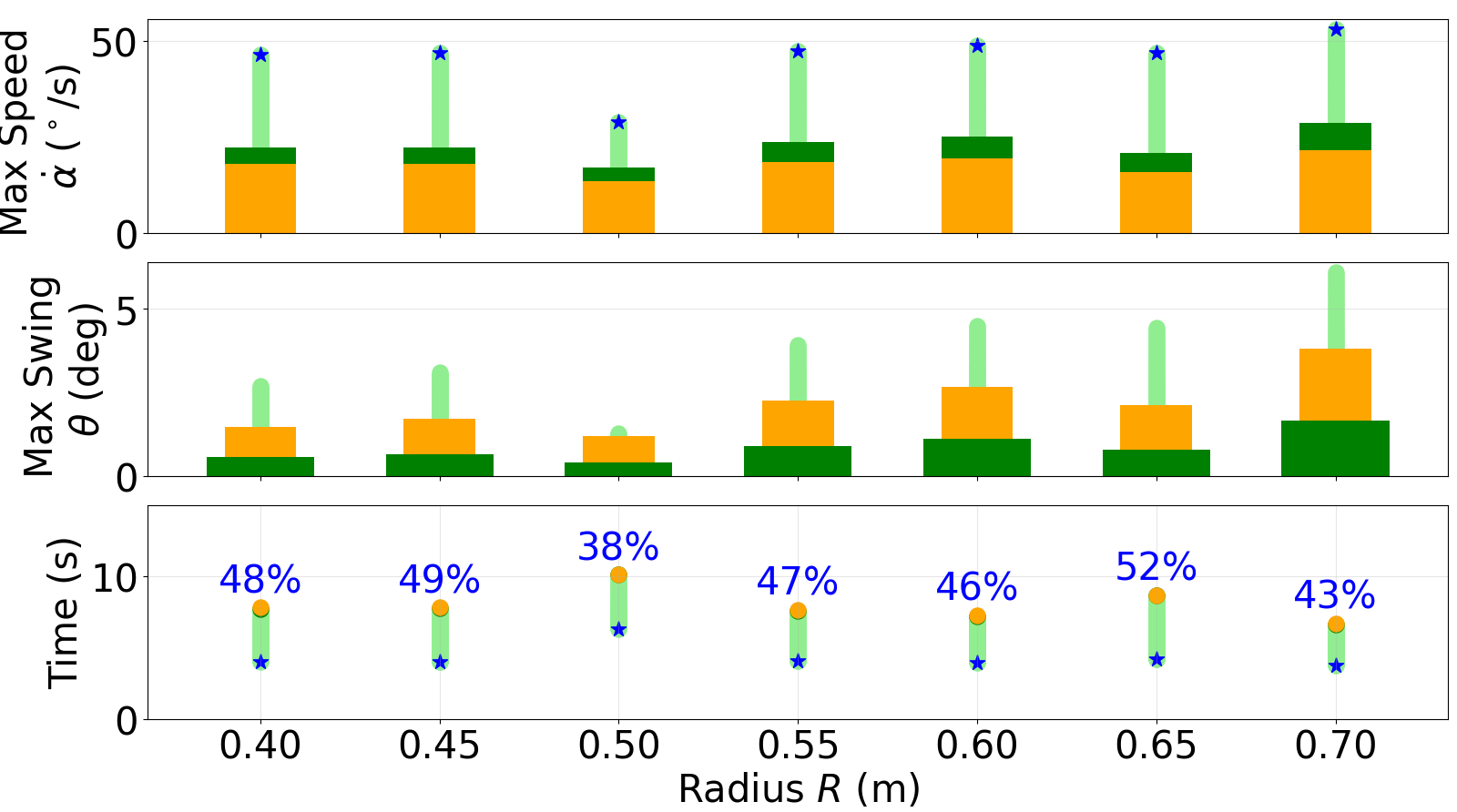}
    \caption{Comparison of slewing performance, with and without input shaping, for each boom radius and load mass $(R,m)$ pair from Fig.~\ref{fig_speed_analysis} at boom length $L_b=0.9144$~m. }
    \label{fig_input_shaping_results}
\end{figure}

The performance of the scaled crane model, with and without input shaping, is compared in Fig.~\ref{fig_input_shaping_results} for each boom radius and payload mass pair $(R,m)$ and a fixed boom length $L_b = 0.9144$~m. In each case, the crane executed a $90\dg$ rotation at increasing slewing speeds. 
The orange bars represent the maximum safe slewing speeds achieved without input shaping. The dark green bars represent cases where shaping allows a higher slewing speed but with a completion time nearly equal to that of the unshaped case with smaller swing angles. The light green markers identify cases where input shaping enables operation at substantially higher speeds (without tip-over), resulting in shorter completion times. The blue star denotes the highest stable speed attained for that $(R,m)$ pair, corresponding to the minimum task duration. Across all cases, this resulted in at least $38\%$ reduction in completion time compared to the unshaped case. Although peak swing increases in the high-speed (light green and star) cases, the crane does not tip-over. Thus, input shaping increases the feasible slewing speed while reducing task duration without compromising safety.

The results indicate that input shaping expands the safe operating range by increasing the maximum speed at which the crane remains stable. Residual swing is suppressed even during fast motion, preventing the overturning moment from exceeding its limit. Thus, shaped acceleration commands enable faster slewing while maintaining stability, improving both safety margin and task efficiency.

\subsection{Experimental Validation}
\begin{figure}[t]
    \centering
    \includegraphics[width=6.4cm]{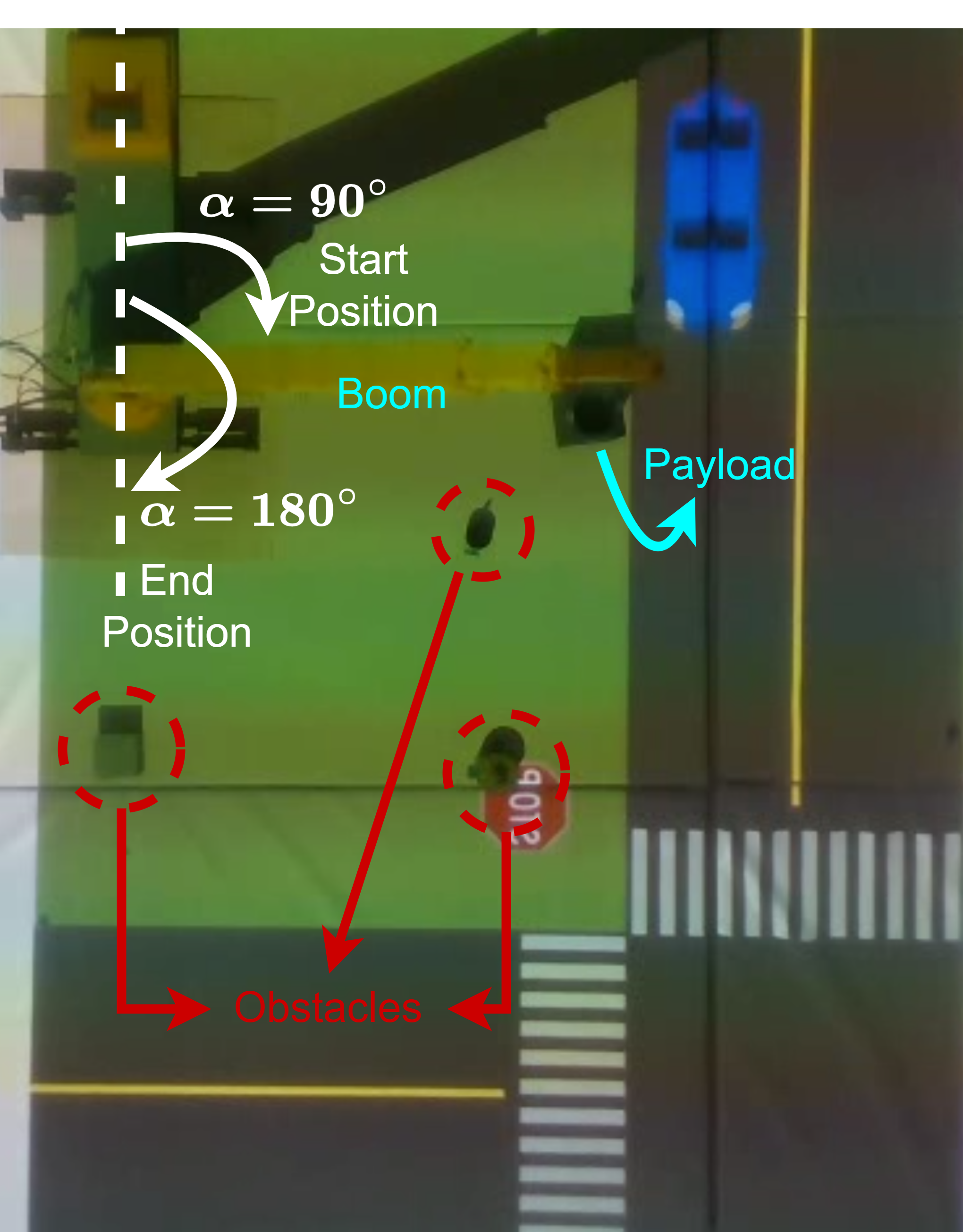}
    \caption{Experimental setup used for human-subject trials. The crane begins at slew angle $\alpha = 90\dg$ and performs a $90\dg$ clockwise slew to $\alpha = 180\dg$ while carrying a suspended payload. Obstacles placed along the trajectory create a constrained workspace in which participants must avoid collisions and prevent tip-over.}
    \label{fig_exp_setup}
\end{figure}
\begin{figure}[b]
    \centering
    \includegraphics[width=8.4cm]{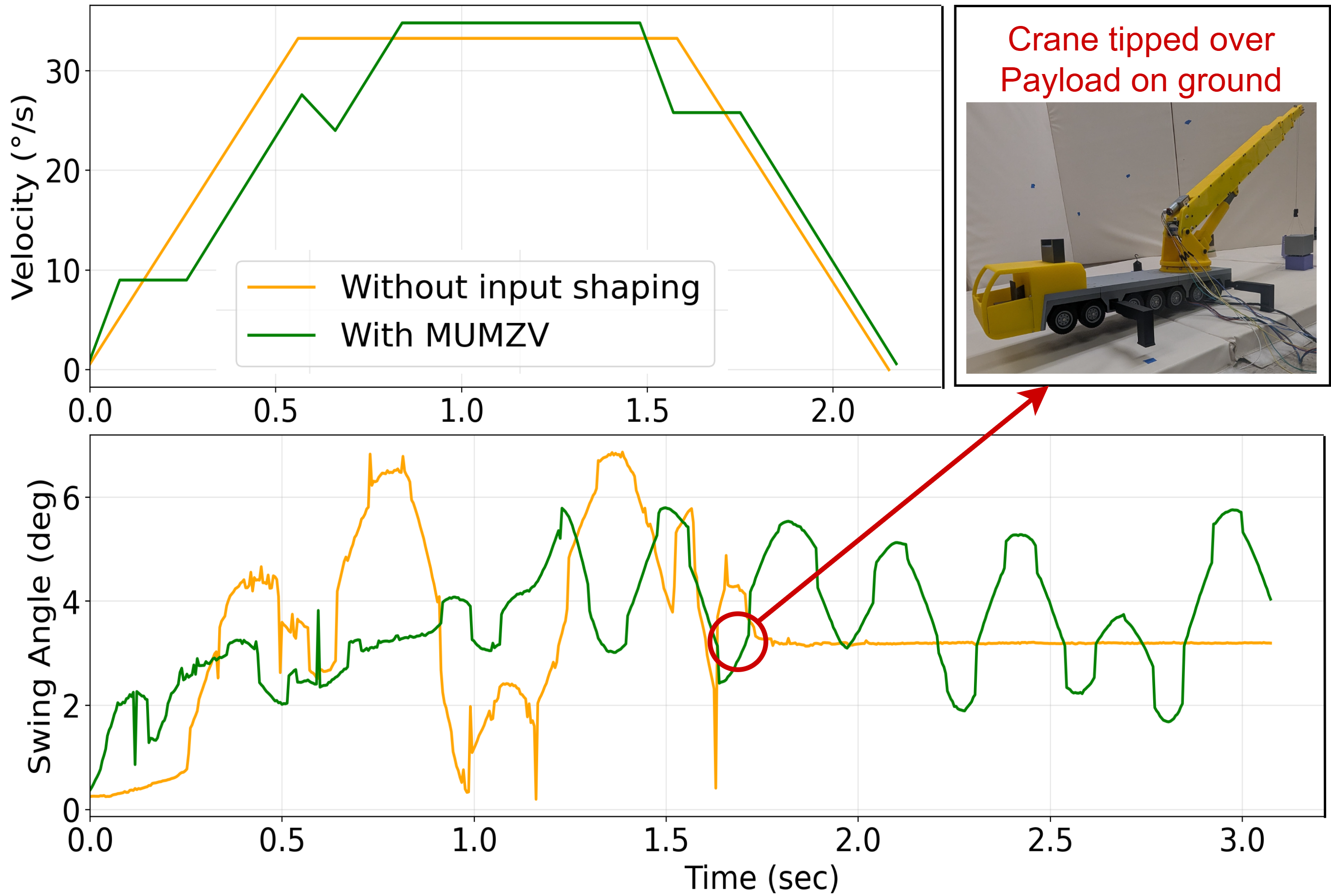}
    \caption{Comparison of unshaped and MUMZV-shaped slewing during a controlled $90\dg$ experiment. Top: Commanded slewing velocity profiles for the unshaped (orange) and MUMZV-shaped (green) cases. Bottom: Measured swing-angle response. }
    \label{fig_auto_experiment}
\end{figure}
\begin{figure*}[ht!]
    \centering
    \includegraphics[width = \linewidth]{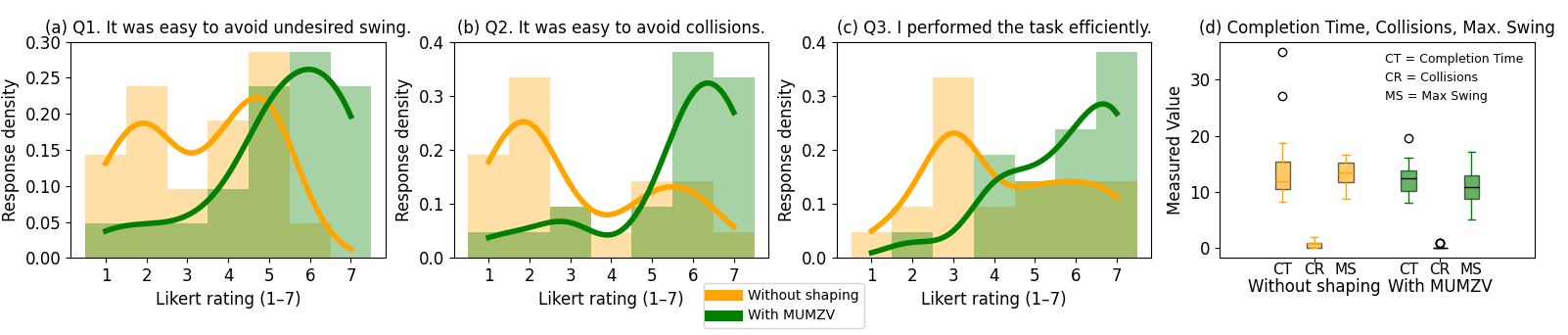}
    \caption{Results from 42 participant-operated $90\dg$ slewing trials performed with and without input shaping. (a-c) Histogram of response density of the subjective questionnaire with seven-point Likert scale comparing responses collected with and without input shaping. The general trend of response density is shown with a smoothed trend curve generated using a squared-exponential kernel with bandwidth 0.7. (d) Quantitative outcome measures, including maximum swing angle (MS), number of collisions (CR), and completion time (CT) for both control modes. }
    \label{fig_exp_results}
\end{figure*}
\subsubsection{Experimental Setup}
A physical experiment was conducted using the scaled mobile crane model shown in Fig.~\ref{fig_sim_real_model}. The boom length $L_b=0.9144$~m and operating radius $R=0.7$~m were fixed, and the payload mass was set to $m=0.5$~kg. The suspended rope length was $l=0.5715$~m. The crane was placed in an obstacle course designed to represent a realistic operating environment. A video of two cars crossing nearby was used (\cite{Pixabay_StopSign_2016}) to simulate roadside traffic, and additional obstacles were placed around the crane to create potential collision hazards as shown in Fig.~\ref{fig_exp_setup}. The experimental platform uses the same geometric dimensions as the simulation model. The payload was chosen to be smaller than the statically stable payload to illustrate the potential for tip-over due to swing.

\subsubsection{Automated Experiment Results}
A controlled $90\dg$ slewing experiment was conducted to compare swing behavior and stability with and without MUMZV shaping. Fig.~\ref{fig_auto_experiment} shows the commanded slewing velocity (top plot) and the corresponding swing response (bottom plot). Although the shaped case (green) operates at a higher slewing speed than the unshaped case (orange), it maintains bounded oscillations throughout the slewing motion. In contrast, the unshaped trial exhibits increasing oscillations that grow rapidly during the motion and ultimately cause the crane to tip-over (annotated in the plot with an accompanying photograph).  Since the payload is on the ground, the swing angle becomes constant after tip-over. Despite the higher velocity, the shaped trial completes the full $90\dg$ rotation without instability, demonstrating that MUMZV shaping effectively suppresses swing growth and prevents swing-induced tip-over.

\subsubsection{Human Trial Results}
The proposed approach was evaluated for swing suppression during manual operation, where variability in operator input can introduce additional swing excitation. Participants were instructed to complete a $90\dg$ slewing motion as fast as possible while avoiding contact with surrounding obstacles and preventing tip-over. Each participant performed the task 6 times, randomly choosing between unshaped manual control and MUMZV-shaped input. After each trial, participants answered the following Likert-scale questions (1 = strongly disagree, 4 = neutral, 7 = strongly agree):
\begin{enumerate}[label=Q\arabic*.]
    \item It was easy to avoid undesired swing.
    \item It was easy to avoid collisions.
    \item I performed the task efficiently.
\end{enumerate}

Quantitative performance metrics were recorded for all trials, including completion time and peak swing angle. A trial was considered failed if the crane tipped over at any point, in which case the participant was required to restart, and the results are presented in  Fig.~\ref{fig_exp_results}:

(1) Reduced swing: Participants reported a decrease in undesired swing when using input shaping compared to the case without input shaping, as shown through responses to Q1 in Fig.~\ref{fig_exp_results}(a). This observation is supported by a Wilcoxon signed-rank test, which revealed a consistent difference between the two conditions, rejecting the null hypothesis $(p<0.05)$. The measured maximum swing values (MS) in Fig.~\ref{fig_exp_results}(d) reflect the same trend, with shaped trials producing smaller peak oscillations. All shaped trials were completed without tip-over, whereas unshaped trials occasionally produced large swings that led to tip-over. Input shaping reduced the average peak swing from $13.1\dg$ to $10.7\dg$, an $18\%$ reduction, with all shaped runs completing without tip-over.

(2) Reduced collisions: Participants reported fewer collision events when using input shaping compared to the case without input shaping, as shown through responses to Q2 in Fig.~\ref{fig_exp_results}(b). This observation is supported by a Wilcoxon signed-rank test, which revealed a consistent difference between the two conditions, rejecting the null hypothesis $(p<0.05)$. The collision counts (CR) in Fig.~\ref{fig_exp_results}(d) are consistent with these responses, with unshaped trials resulting in more obstacle contacts while shaped trials were largely collision-free. All shaped trials were completed without tip-over, indicating improved operational safety around obstacles.  Collision events decreased from an average of 0.80 per trial to 0.14, an $82\%$ reduction.

(3) Improved efficiency: Participants reported higher perceived task efficiency when input shaping was applied, even without considering failure cases such as tip-over that required restarts, as shown in the Q3 responses in Fig.~\ref{fig_exp_results}(c). This observation is supported by a Wilcoxon signed-rank test, which indicated a consistent difference between the two conditions, rejecting the null hypothesis $(p<0.05)$. The completion-time measurements (CT) in Fig.~\ref{fig_exp_results}(d) are consistent with these responses, with shaped runs completing faster on average. All shaped trials completed successfully, demonstrating that improved efficiency was achieved without compromising stability. Shaped trials completed faster, reducing time from $13.8$~s to $12$~s, $13\%$ decrease, while remaining stable.

\section{Conclusion}

This work demonstrated that input shaping can increase the safe operating speed of mobile cranes while maintaining stability against tip-over. By shaping the slewing acceleration to avoid exciting the payload swing mode, the crane remained stable at higher speeds, and the slewing task was completed in substantially less time compared to unshaped control. Simulation analysis and  physical scaled-crane experiments both confirmed that input shaping expanded the safe speed envelope, suppressed residual oscillations, and reduced the likelihood of swing-induced tip-over. User trials further indicated improved perceived efficiency and ease of operation when input shaping was applied.

Future work will extend the input-shaping approach to conditions with varying rope length and changing boom elevation, where the natural swing frequency is no longer fixed. Additional evaluation under wind disturbances will help assess robustness in outdoor scenarios. Finally, applying the method to a full-scale mobile crane will enable validation under real operational loads and environmental variability.

\begin{ack}
Support for this work was provided by the Sarah Pantip Wong Family through the Crane Safety Research Center at Georgia Tech and the University of Washington. The authors thank Nathaneal Bursch, Alex Cong, Joe Duffie, Zach Ridgway, and Astro Trebach for  designing and building the scaled mobile crane used in the experiments for this paper. 
\end{ack}

\section*{DECLARATION OF GENERATIVE AI AND AI-ASSISTED TECHNOLOGIES IN THE WRITING PROCESS}
During the preparation of this work the author(s) used OpenAI ChatGPT in order to support debugging and improve grammatical clarity. After using this tool, the author(s) reviewed and edited the content as needed and take(s) full responsibility for the content of the publication.

\bibliography{cite}
\end{document}